\documentclass[12pt]{article}
 \usepackage{epsfig}
 \def\be{\begin{equation}}
 \def\ee{\end{equation}}
 \def\bea{\begin{eqnarray}}
 \def\eea{\end{eqnarray}}
 \usepackage{graphicx}

 \catcode`\@=11
 \def\lsim{\mathrel{\mathpalette\@versim<}}
 \def\gsim{\mathrel{\mathpalette\@versim>}}
 \def\@versim#1#2{\vcenter{\offinterlineskip
 \ialign{$\m@th#1\hfil##\hfil$\crcr#2\crcr\sim\crcr } }}
 \catcode`\@=12

 \catcode`@=12
\begin{document}
 \thispagestyle{empty}
 \begin{flushright}
 UCRHEP-T572\\
 November 2016\
 \end{flushright}
 \vspace{0.8in}
 \begin{center}
 {\LARGE \bf Pathways to Naturally Small\\ Dirac Neutrino Masses\\}
 \vspace{1.5in}
 {\bf Ernest Ma and Oleg Popov\\}
 \vspace{0.2in}
 {\sl Department of Physics and Astronomy,\\ 
 University of California, Riverside, California 92521, USA\\}
 \end{center}
 \vspace{1.5in}

\begin{abstract}\
If neutrinos are truly Dirac fermions, the smallness of their masses 
may still be natural if certain symmetries exist beyond those of the 
standard model of quarks and leptons.  We perform a systematic study of 
how this may occur at tree level and in one loop.  We also propose a 
scotogenic version of the left-right gauge model with naturally small 
Dirac neutrino masses in one loop. 
\end{abstract}

\newpage
 \baselineskip 24pt

\noindent \underline{\it Introduction}~:\\
If neutrinos are Majorana fermions, then it has been known~\cite{w79} 
since 1979 that they are described by a unique dimension-five operator 
beyond the standard model of quarks and leptons, i.e.
\begin{equation}
{\cal L}_5 = -{f_{ij} \over 2 \Lambda} (\nu_i \phi^0 - l_i \phi^+)
(\nu_j \phi^0 - l_j  \phi^+) + H.c.
\end{equation}
Neutrino masses are then proportional to $v^2/\Lambda$, where $v = \langle 
\phi^0 \rangle$ is the vacuum expectation value of the Higgs doublet 
$(\phi^+,\phi^0)$.  This formula is necessarily seesaw because $\Lambda$ 
has already been assumed to be much greater than $v$ in the first place. 
It has also been known~\cite{m98} since 1998 that there are three 
specific tree-level realizations (denoted as Types I,II,III) and three 
generic one-particle-irreducuble one-loop realizations.

If neutrinos are Dirac fermions, then the 
term $m_D \bar{\nu}_L \nu_R \bar{\phi}^0$ is desired but 
$(m_N/2) \nu_R \nu_R$ must be forbidden.  This requires 
the existence of a symmetry, usually taken to be global $U(1)_L$ lepton 
number.  This may be the result of a spontaneously broken $U(1)_{B-L}$ 
gauge symmetry where the scalar which breaks the symmetry carries 
three~\cite{ms15} and not two units of $B-L$ charge.  On the other 
hand, global $U(1)_L$ is not the only possibility.  The notion of lepton 
number itself may in fact be discrete.  It cannot of course be $Z_2$, 
then $m_N$ would be allowed and neutrino masses are Majorana.  However, 
it may be $Z_3$~\cite{mpsz15,bmpv16} or $Z_4$~\cite{hr13,h13,cmrv16}, 
but then new particles must appear to legitimize this discrete 
lepton symmetry.  Since there are three neutrinos, a flavor symmetry 
may also be used to forbid the $\nu_R \nu_R$ terms~\cite{abmpv14}.

To obtain a naturally small $m_D$, there must be another symmetry which 
forbids the dimension-four $\bar{\nu}_L \nu_R \bar{\phi}^0$ term, but this 
symmetry must also be softly or spontaneously broken, so that an effective 
$m_D$ appears, at tree level or in one loop, suppressed by large masses.  
The symmetry used to achieve this is model-dependent.  Nevertheless 
generic conclusions may be obtained regarding the nature of the necessary 
particles involved, as shown below.

\newpage
\noindent \underline{\it Four specific tree-level realizations}~:\\
Assume a symmetry ${\cal S}$ under which $\nu_L$ and $\phi^0$ do not transform, 
but $\nu_R$ does.  There are then four and only four ways to connect 
them at tree level through the soft breaking of this symmetry.
\begin{itemize}

\item Insert a Dirac fermion singlet $N$ which does not transform under 
${\cal S}$, then break ${\cal S}$ softly by the dimension-three  
$\bar{\nu}_R N_L$ term.

\begin{figure}[htb]
\vspace*{-4cm}
\hspace*{-3cm}
\includegraphics[scale=1.0]{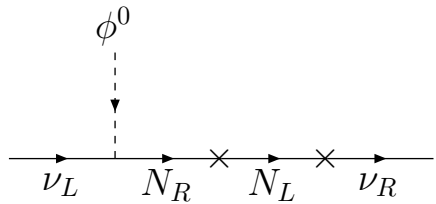}
\vspace*{-21.5cm}
\caption{Dirac neutrino mass with a Dirac singlet fermion insertion.}
\end{figure}

\item Insert a Dirac fermion triplet $(\Sigma^+,\Sigma^0,\Sigma^-)$ which 
does not transform under ${\cal S}$, then break ${\cal S}$ and 
$SU(2)_L \times U(1)$ together spontaneously to obtain the dimension-three 
$\bar{\nu}_R \Sigma^0_L$ term.

\begin{figure}[htb]
\vspace*{-4cm}
\hspace*{-3cm}
\includegraphics[scale=1.0]{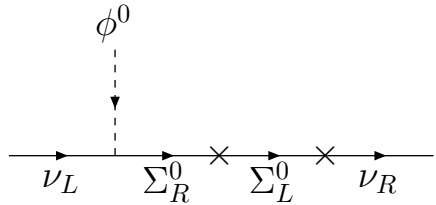}
\vspace*{-21.5cm}
\caption{Dirac neutrino mass with a Dirac triplet fermion insertion.}
\end{figure}

\item Insert a Dirac fermion doublet $(E^0,E^-)$ which transforms as $\nu_R$ 
under ${\cal S}$, then break ${\cal S}$ softly by the dimension-three 
$(\bar{E}^0 \nu_L + E^+ e^-)$ term.

\begin{figure}[htb]
\vspace*{-4cm}
\hspace*{-3cm}
\includegraphics[scale=1.0]{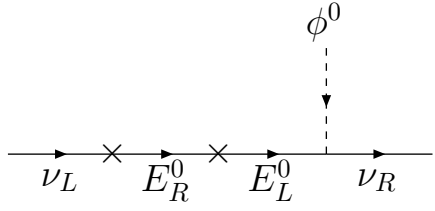}
\vspace*{-21.5cm}
\caption{Dirac neutrino mass with a Dirac doublet fermion insertion.}
\end{figure}

\newpage
\item Insert a scalar doublet $(\eta^+,\eta^0)$ which transforms as $\nu_R$ 
under ${\cal S}$, then break ${\cal S}$ softly by the dimension-two 
$(\eta^- \phi^+ + \bar{\eta}^0 {\phi}^0)$ term.

\begin{figure}[htb]
\vspace*{-3cm}
\hspace*{-3cm}
\includegraphics[scale=1.0]{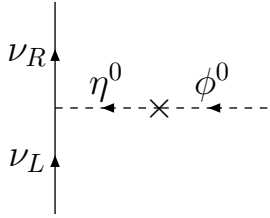}
\vspace*{-21.5cm}
\caption{Dirac neutrino mass with a doublet scalar insertion.}
\end{figure}

\end{itemize}

In Figs.~1 to 3, the mechanism which makes $m_D$ small is the Dirac 
seesaw~\cite{rs84}. 
The $2 \times 2$ mass matrix linking $(\bar{\nu}_L.\bar{\psi}_L)$ to 
$(\nu_R,\psi_R)$, where $\psi = N, \Sigma^0, E^0$, is of the form 
\begin{equation}
{\cal M}_{\nu \psi} = \pmatrix{0 & m_1 \cr m_2 & M_\psi}.
\end{equation}
Since $M_\psi$ is an invariant mass, it may be assumed to be large, 
whereas $m_{1,2}$ come from either electroweak symmetry breaking or 
${\cal S}$ breaking and may be assumed small in comparison. 
Hence $m_D \simeq m_1 m_2/M_\psi$ is naturally small as desired.

In Fig.~4, the mechanism is also seesaw but in the scalar sector, as 
first pointed out in Ref.~\cite{m01}.  Using the small soft ${\cal S}$ 
breaking term $\bar{\eta}^0 \phi^0$ together with a large mass for $\eta$, 
a small vacuum expectation value $\langle \eta^0 \rangle$ is induced to 
obtain $m_D$~\cite{dl09}.  This may also be accomplished by extending the 
gauge symmetry~\cite{vv16,rvv16}.

\noindent \underline{\it Two generic one-loop realizations}~:\\
Suppose the new particles considered previously for connecting $\nu_L$ 
with $\nu_R$ at tree level are not available, then a Dirac neutrino mass 
may still occur in one loop.  Assuming that this loop consists of a 
fermion line and a scalar line, then the external Higgs boson must 
couple to either the scalar line or the fermion line, yielding two 
generic diagrams.

\begin{itemize}

\item Consider the one-loop connection shown below.
\begin{figure}[htb]
\vspace*{-3cm}
\hspace*{-3cm}
\includegraphics[scale=1.0]{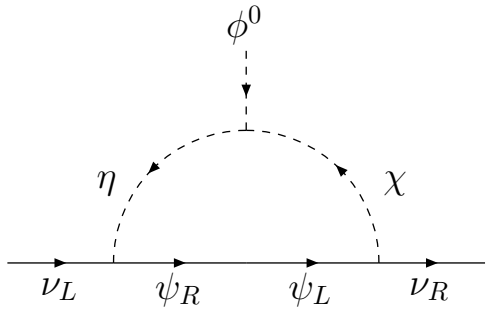}
\vspace*{-21.5cm}
\caption{Dirac neutrino mass in one loop with trilinear scalar coupling.}
\end{figure}
Since $\nu_R$ transforms under ${\cal S}$ and $\nu_L$ and $\phi^0$ do 
not, a Dirac neutrino mass is only generated if ${\cal S}$ is broken 
softly by either the dimension-three $\bar{\psi}_L \psi_R$ term or 
the dimension-three $\bar{\eta} \chi \phi^0$ term.  There are an 
infinite number of solutions for the new fermion $\psi$ and the new 
scalars $\eta$ and $\chi$.  Under the electroweak $SU(2)_L \times U(1)_Y$, 
the three simplest solutions are listed in Table 1.
\begin{table}[htb]
\caption{$SU(2)_L \times U(1)_Y$ assigments of $\psi$, $\eta$, and $\chi$.}
\begin{center}
\begin{tabular}{|c|c|c|c|}
\hline
solution & $\psi$ & $\eta$ & $\chi$ \\
\hline
A & (1,0) & $(2,-1/2)$ & (1,0) \\
\hline
B & (2,1/2) & (1,0) & (2,1/2) \\
\hline
C & $(2,-1/2)$ & $(1,-1)$ & $(2,-1/2)$ \\
\hline
\end{tabular}
\end{center}
\end{table}
Note that solutions also exist with $\psi,\chi,\eta$ all carrying color. 
Let ${\cal S}$ be $Z_2$ as an example, then the assignments of $\eta$, 
$\psi_R$, $\psi_L$, and $\chi$ under ${\cal S}$ are given in Table 2.
\begin{table}[htb]
\caption{${\cal S}=Z_2$ assigments of $\eta$, $\psi_R$, $\psi_L$, and $\chi$.}
\begin{center}
\begin{tabular}{|c|c|c|c|c|c|c|}
\hline
solution & $\eta$ & $\psi_R$ & $\psi_L$ & $\chi$ & $\bar{\psi}_L \psi_R$ 
& $\bar{\eta} \chi \phi^0$ \\
\hline
A1 & $-$ & $-$ & $-$ & + & + & $-$ \\
A2 & $-$ & $-$ & $+$ & $-$ & $-$ & $+$ \\
\hline
B1 & $+$ & $+$ & $+$ & $-$ & + & $-$ \\
B2 & $-$ & $-$ & $+$ & $-$ & $-$ & $+$ \\
\hline
C1 & $+$ & $+$ & $+$ & $-$ & + & $-$ \\
C2 & $-$ & $-$ & $+$ & $-$ & $-$ & $+$ \\
\hline
\end{tabular}
\end{center}
\end{table}
The solutions A1 and B1 must be discarded, because $\chi$ and $\eta$ are 
neutral scalar singlets which are even under $Z_2$ respectively.  As such, 
they will acquire vacuum expectation values from interactions with $\Phi$. 
From the trilinear coupling $\bar{\eta}\chi\phi^0$, this in turn would 
induce a vacuum expectation value for $\eta$ and $\chi$ in A1 and B1 
respectively.  Hence the loop of Fig.~5 would collapse to a tree as 
shown in Figs.~1 and 3.  The solutions A2(B2) should also be discarded 
because $\psi_{R,L}$ transform exactly as $\nu_{R,L}$, thus collapsing 
to Fig.~4.  However, these solutions could be reinstated with the 
scotogenic mechanism to be discussed later.

\item Consider now the other possible connection.
\begin{figure}[htb]
\vspace*{-3cm}
\hspace*{-3cm}
\includegraphics[scale=1.0]{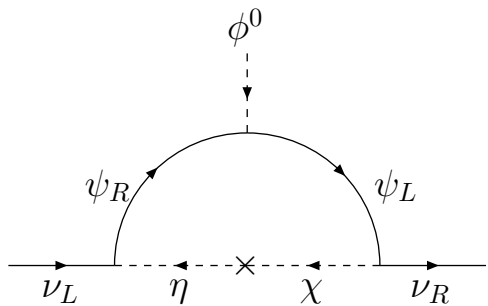}
\vspace*{-21.5cm}
\caption{Dirac neutrino mass in one loop with quadratic scalar mixing.}
\end{figure}
The only soft term here is the quadratic $\bar{\eta} \chi$ term which must 
be odd under ${\cal S}=Z_2$.  If $\eta$ and $\chi$ are neutral, then again 
they must have vacuum expectation values, thus collapsing the loop of 
Fig.~6 to a tree.  Hence $\eta$ and $\chi$ must be charged or colored, 
and if $\chi \sim \pm$ under ${\cal S}$, then $\psi_{L,R}, \eta \sim \mp$.
An example of such a model is Ref.~\cite{kns11}.  It has also been 
implemented in left-right gauge models many years ago~\cite{m87,bm89}.

\end{itemize}

In the above, there must be of course also a symmetry which maintains 
lepton number.  This symmetry may propagate along the fermion line in the 
loop, which is the conventional choice, but it may also propagate along 
the scalar line in the loop.  If the latter, then lepton number may serve 
as the stabilizing symmetry of dark matter~\cite{m15}.  The reason is 
very simple.  For the lightest scalar, say $\eta$, having lepton number 
which is conserved, it can only decay into a lepton plus a fermion which has 
no lepton number, say $\psi$, and vice versa.  Hence the lightest $\psi$ or 
the lightest $\eta$ is dark matter.  This means that the loop diagrams 
of Figs.~5 and 6 could be naturally scotogenic, from the Greek 'scotos' 
meaning darkness.  This mechanism was invented 10 years ago~\cite{m06}.  The 
unconventional assignment of lepton number to scalars and fermions also 
reinstates the solutions A,B considered earlier, because now the particles 
in the loop have odd dark parity, as discussed in Ref.~\cite{gs08,fm12}.  
This application of the one-loop diagram for Dirac neutrino mass 
using scalars carrying lepton number is actually well-known in 
supersymmetry, where the exchange of sleptons and neutralinos 
contributes to charged-lepton masses.  Here we show that the generic 
idea is also applicable without supersymmetry. 

\noindent \underline{\it Scotogenic Dirac neutrino mass in left-right model}~:\\
The absence of a tree-level Dirac neutrino mass may be due to the 
underlying gauge symmetry and the scalar particle content. 
Consider the following left-right gauge mode based on 
$SU(3)_C \times SU(2)_L \times SU(2)_R \times U(1)_X$ together with a 
discrete $Z_2$ symmetry.  
It extends the standard model (SM) to include heavy charged quarks and 
leptons which are odd under $Z_2$, but no scalar bidoublet~\cite{bms03} as 
shown in Table 3.  Its fermion content is identical to a recent 
proposal~\cite{bkmtz16} at this point.
\begin{table}[htb]
\caption{Particle content of proposed left-right gauge model.}
\begin{center}
\begin{tabular}{|c|c|c|c|c|c|}
\hline
particles & $SU(3)_C$ & $SU(2)_L$ & $SU(2)_R$ & $U(1)_X$ & $Z_2$ \\
\hline
$(u,d)_L$ & 3 & 2 & 1 & 1/6 & + \\
$u_{R}$ & $3$ & 1 & 1 & 2/3 & + \\
$d_{R}$ & $3$ & 1 & 1 & $-1/3$ & + \\
$(\nu,e)_L$ & 1 & 2 & 1 & $-1/2$ & + \\
$e_{R}$ & 1 & 1 & 1 & $-1$ & $+$ \\
\hline
$(U,D)_R$ & 3 & 1 & 2 & 1/6 & $-$ \\
$(\nu,E)_R$ & 1 & 1 & 2 & $-1/2$ & $-$ \\
$U_{L}$ & $3$ & 1 & 1 & 2/3 & $-$ \\
$D_{L}$ & $3$ & 1 & 1 & $-1/3$ & $-$ \\
$E_{L}$ & 1 & 1 & 1 & $-1$ & $-$ \\
\hline
$(\phi_L^+,\phi_L^0)$ & 1 & 2 & 1 & $1/2$ & + \\
$(\phi_R^+,\phi_R^0)$ & 1 & 1 & 2 & $1/2$ & + \\
\hline
\end{tabular}
\end{center}
\end{table}

\begin{table}[htb]
\caption{Scotogenic additions to the proposed left-right gauge model.}
\begin{center}
\begin{tabular}{|c|c|c|c|c|c|c|}
\hline
particles & $SU(3)_C$ & $SU(2)_L$ & $SU(2)_R$ & $U(1)_X$ & $Z_2$ & $Z_2^D$ \\
\hline
$N_{L,R}$ & 1 & 1 & 1 & 0 & + & $-$ \\
\hline
$(\eta_L^+,\eta_L^0)$ & 1 & 2 & 1 & $1/2$ & + & $-$ \\
$(\eta_R^+,\eta_R^0)$ & 1 & 1 & 2 & $1/2$ & $-$ & $-$ \\
$\chi_L$ & 1 & 1 & 1 & 0 & + & $-$ \\
$\chi_R$ & 1 & 1 & 1 & 0 & $-$ & $-$ \\
$\chi_0$ & 1 & 1 & 1 & 0 & $-$ & $+$ \\
\hline
\end{tabular}
\end{center}
\end{table}
The breaking of $SU(2)_{L,R}$ is accomplished by the Higgs doublets 
$\Phi_{L,R}$. 
The SM quarks and charged leptons obtain masses from $\Phi_L$.  The heavy 
quarks and charged leptons obtain masses from $\Phi_R$.  They are separated 
by the $Z_2$ symmetry and do not mix.  Both $\nu_L$ and $\nu_R$ are 
massless and separated by $Z_2$.  To link them with a Dirac mass, this 
$Z_2$ has to be broken.  This is implemented as shown in Table 4 using 
the unbroken symmetry $Z_2^D$ for dark matter, under which $N, \eta_{L,R}, 
\chi_{L,R}$ are odd and all others are even.  The symmetry $Z_2$ is assumed to 
be respected by all dimension-three terms as well, so there is no 
$\bar{Q}_L q_R$ or $\bar{E}_L e_R$ term.  Hence $V_{CKM}$ remains unitary 
as in the SM.  It is broken only by 
the unique dimension-two term $\chi_L \chi_R$.  The resulting scotogenic 
diagram for Dirac neutrino mass is shown in Fig.~7.
\begin{figure}[htb]
\vspace*{-3cm}
\hspace*{-3cm}
\includegraphics[scale=1.0]{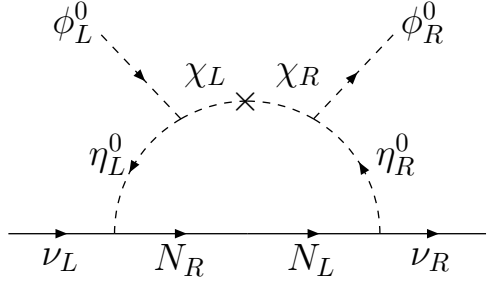}
\vspace*{-21.5cm}
\caption{Scotogenic Dirac neutrino mass in left-right symmetry.}
\end{figure}
The connection between the heavy fermions of the $SU(2)_R$ sector and 
the SM fermions is $\chi_0$ with the allowed dimension-four Yukawa couplings 
$\chi_0 \bar{Q}_L q_R$ and $\chi_0 \bar{E}_L e_R$.  Now $\chi_0$ mixes 
only radiatively with the SM Higgs boson, a phenomenon discovered only 
recently~\cite{m16}, and decays to SM particles but its lifetime may be 
long.  At the Large Hadron Collider, the heavy $SU(2)_R$ quarks are 
easily produced if kinematically allowed.  The lightest will decay to 
a SM quark and $\chi_0$ which may escape the detector as missing energy. 
This has the same signature as dark matter.  The true dark matter is of 
course the lightest neutral fermion or boson with odd $Z_2^D$. 

\noindent \underline{\it Concluding remarks}~:\\ 
The notion that neutrino masses are Dirac is still viable in the absence 
of incontrovertible experimental proof of the existence of neutrinoless 
double beta decay.  The theoretical challenge is to understand why.  
In this paper we study systematically how the smallness of Dirac neutrino 
masses may be achieved at tree level (four specific cases) and in one loop 
(two generic cases).  We also propose a scotogenic left-right gauge model.

\noindent \underline{\it Acknowledgement}~:\\
This work was supported in part by the U.~S.~Department of Energy Grant 
No. DE-SC0008541.

\bibliographystyle{unsrt}

\begin{thebibliography}{99}
\bibitem{w79} S. Weinberg, Phys. Rev. Lett. {\bf 43}, 1566 (1979).
\bibitem{m98} E. Ma, Phys. Rev. Lett. {\bf 81}, 1171 (1998).
\bibitem{ms15} E. Ma and R. Srivastava, Phys. Lett. {\bf B741}, 217 (2015).
\bibitem{mpsz15} E. Ma, N. Pollard, R. Srivastava, and M. Zakeri, Phys. Lett. 
{\bf B750}, 135 (2015).
\bibitem{bmpv16} C. Bonilla, E. Ma, E. Peinado, and J.~W.~F. Valle, 
Phys. Lett. {\bf B762}, 214 (2016).. 
\bibitem{hr13} J. Heeck and W. Rodejohann, Europhys. Lett. {\bf 103}, 32001 
(2013).
\bibitem{h13} J. Heeck, Phys. Rev. {\bf D88}, 076004 (2013).
\bibitem{cmrv16} S. Centelles Chulia, E. Ma, R. Srivastava, and J.~W.~F. 
Valle, arXiv:1606.04543 [hep-ph].
\bibitem{abmpv14} A. Aranda, C. Bonilla, S. Morisi, E. Peinado, and 
J. W. F. Valle, Phys. Rev. {\bf D89}, 033001 (2014).
\bibitem{rs84} P. Roy and O. U. Shanker, Phys. Rev. Lett. {\bf 52}, 713 (1984).
\bibitem{m01} E. Ma, Phys. Rev. Lett. {\bf 86}, 2502 (2001).
\bibitem{dl09} S. M. Davidson and H. E. Logan, Phys. Rev. {\bf D80}, 095008 
(2009).
\bibitem{vv16} J. W. F. Valle and C. A. Vaquera-Araujo, Phys. Lett. 
{\bf B755}, 363 (2016).
\bibitem{rvv16} M. Reig, J. W. F. Valle, and C. A. Vaquero-Araujo, Phys. 
Rev. {\bf D94}, 033012 (2016).
\bibitem{kns11} S. Kanemura, T. Nabeshima, and H. Sugiyama, Phys. Lett. 
{\bf B703}, 66 (2011).
\bibitem{m87} R. N. Mohapatra, Phys. Lett. {\bf B198}, 69 (1987).
\bibitem{bm89} B. S. Balakrishna and R. N. Mohapatra, Phys. Lett. {\bf B216}, 
349 (1989).
\bibitem{m15} E. Ma, Phys. Rev. Lett. {\bf 115}, 011801 (2015).
\bibitem{m06} E. Ma, Phys. Rev. {\bf D73}, 077301 (2006).
\bibitem{gs08} P.-H. Gu and U. Sarkar, Phys. Rev. {\bf D77}, 105031 (2008).
\bibitem{fm12} Y. Farzan and E. Ma, Phys. Rev. {\bf D86}, 033007 (2012).
\bibitem{bms03} B. Brahmachari, E. Ma, and U. Sarkar, Phys. Rev. Lett. 
{\bf 91}, 011801 (2003).
\bibitem{bkmtz16} P. S. Bhupal Dev, D. Kazanas, R. N. Mohapatra, V. L. 
Teplitz, and Y. Zhang, JCAP {\bf 1608}, 034 (2016).
\bibitem{m16} E. Ma, Phys. Lett. {\bf B754}, 114 (2016).
\end{thebibliography}

\end{document}